\documentclass[12pt]{article}
\usepackage{latexsym}
\usepackage{epsfig,amssymb,euscript}
\usepackage{amsmath}
\usepackage{array,calc,epsfig}

\def\be{\begin{equation}}
\def\ee{\end{equation}}
\def\bea{\begin{eqnarray}}
\def\eea{\end{eqnarray}}

\def\R{{\cal R}}

\begin{document}
\baselineskip=15.5pt
\setcounter{page}{1}
\begin{titlepage}

\rightline{\small{\tt UB-ECM-PF-07/13 }}
\rightline{\small{\tt CERN-PH-TH/2007-093}}

\begin{center}

\vskip 1.4 cm

{\LARGE{\bf Notes on a SQCD-like plasma dual and holographic
renormalization}}

\vskip 1.5cm

{\Large{
A. L. Cotrone $^a$, J.M. Pons $^a$ and P. Talavera $^b$}}\\

\vskip 1.2cm

\textit{$^a$ Departament ECM, Facultat de F\'isica, Universitat de Barcelona and \\ Institut
de Fisica d'Altes Energies, Diagonal 647, E-08028 Barcelona, Spain.}\\
\textit{$^b$ Theory Group, Physics Department, CERN, Geneva 23,
CH-1211 Switzerland.}

\end{center}

\vspace{12pt}

\begin{center}
\textbf{Abstract}
\end{center}

\vspace{4pt}{\noindent We study the thermodynamics and the jet
quenching parameter of a black hole solution dual to a SQCD-like
plasma which includes the backreaction of fundamental flavors. The
free energy is calculated in several ways, including some recently
proposed holographic renormalization prescriptions. The validity of
the latter is confirmed by the consistency with the other methods.
The resulting thermodynamic properties are similar to the Little
String Theory ones: the temperature is fixed at the Hagedorn value
and the free energy is vanishing. Finally, an accurate analysis of
the relevant string configurations shows that the jet quenching
parameter is zero in this model, in agreement with previous
findings. }

\vfill \vskip 5.mm \hrule width 5.cm \vskip 2.mm {\small \noindent e-mail: cotrone@ecm.ub.es, pons@ecm.ub.es, pedro.talavera@cern.ch}

\noindent
\end{titlepage}

\newpage
\section{Introduction and conclusion}

In the context of the string/gauge theory correspondence
\cite{malda}, the study of theories at finite temperature is
recently attracting renewed interest. In fact, from the first works
on the subject \cite{klebanov,WittenZW}, the investigation of properties of
thermal field theories and the thermodynamics of the dual black
holes have revealed an exciting correspondence between these two
previously independent topics. In particular, the calculation of
quantities such as the free energy, entropy or energy density can be
performed in the gravity side, providing results for the
thermodynamics of the dual field theories. Since in performing such
computations the observables typically turn out to be infinite, one
has to employ renormalization techniques in order to extract the
meaningful physical informations. One such procedure, which goes
under the name of ``holographic renormalization'', is particularly
interesting, since it provides a way of renormalizing infinities in
gravity computations, under the requirement that the dual
observables in field theory must be finite
\cite{Balasubramanian:1999re,deHaro:2000xn,KarchMS}, thus connecting
the counterterm renormalization in field theory with a specific
subtraction mechanism on the gravity side.

Moreover, recently the string/gauge theory correspondence for
thermal systems has become of great importance because of the
evidence that string theory may provide a useful tool to study some
properties of Quark Gluon Plasmas (QGPs) as the one experimentally
produced at RHIC \cite{rhic}. Relevant examples include hydrodynamic
features, jet quenching phenomena, screening properties and
photoproduction (see \cite{pss}-\cite{liu3} and the references to
these works).

Since the string models usually employed for the latter task do not
account for the backreaction of fundamental flavors, with the
exception of \cite{noi}, it is clearly of interest to try and study
what are the effects of the ``quarks'' on the stringy predictions
for the QGP. In this note we consider the only known ten dimensional
black hole solution dual to a four dimensional, finite temperature
field theory that includes the backreaction of many flavor degrees
of freedom \cite{CaseroPT}. The zero temperature solution is dual to
a ${\cal N}=1$ SQCD theory with a superpotential, with number of
flavors $N_f$ being the double of the number of colors $N_c$,
$N_f=2N_c$. The solution corresponds to wrapped ``color'' D5-branes
and smeared ``flavor'' D5-branes.

After the introduction of the solution in section \ref{introbh}, we
study its thermodynamic properties in section \ref{termodyn}. The
results are very similar to the Little String Theory ones,
consistently with the fact that the solution comes from D5-branes.
The temperature of the black hole is fixed at the Hagedorn value and
it is independent on the horizon size. Thus, the free energy density
turns out to be exactly zero in the gravity approximation: the
energy density is equal to the entropy density times the
temperature.

We perform such analysis in three ways. As a first route, we
renormalize the infinities in the gravity computation of the free
energy by subtracting the analogous contributions of a reference
background, namely the zero temperature solution. Alternatively, we
separately calculate the energy (by subtracting the reference
background value) and the entropy density. Finally, we employ very
simple holographic renormalization prescriptions using two different
counterterms, one for the ordinary gravity plus surface terms and
one for the flavor terms. While the counterterms for the gravity
plus surface contributions are fairly standard, the ones for the
flavor terms are in principle on a less solid ground, since they
have been much less tested.

In fact, in the model at hand the flavor contribution is included by
sourcing the supergravity fields with the Born-Infeld action for the
smeared D5-branes. In the literature there exist formulas for the
counter-terms for {\emph{probe}} flavor branes
\cite{KarchMS,SkenderisUY,MateosVN}. We show that the latter are
sufficient to account for the renormalization of the flavor
contribution also in the case where the full backreaction of the
flavor branes is taken into account. The latter result is not
a-priori guaranteed, and the situation could
have also been worsened by the smearing procedure. Instead, we
find that the using the two counterterms proposed in \cite{MateosVN}
gives a completely consistent picture of the thermodynamics of the
backreacted system.
Let us also point out that these two simple
counterterms give in the present case a unique renormalization
prescription, thus providing a complementary approach to holographic
renormalization to the one employed for example in \cite{MarolfYS}
for the Little String Theory case (we briefly review the LST case in
our setting in section \ref{seclst}).

Finally, in a rather independent part of the paper (section
\ref{quenching}), we come to the issue of the evaluation of the jet
quenching parameter $\hat q$. The latter was computed in the model at
hand in \cite{noi} using the string configuration proposed in
\cite{LiuUG}, with an unexpected vanishing result, $\hat q$=0. Given
this behavior and considering the proposal made in \cite{argy} of
using a different string solution for the calculation of $\hat q$,
we perform an accurate analysis of the configurations that could be
relevant for the purpose.\footnote{We thank F. Bigazzi for pointing
out the possible relevance of this analysis.} We find indications
that the only solution that gives a meaningful result (for the
action employed for the calculation of $\hat q$) gives a vanishing
jet quenching parameter, in agreement with the finding in
\cite{noi}.

\section{Introducing the black hole}\label{introbh}
We shall deal with a background  which is the thermal deformation of
the dual to a ${\cal N}=1$ SQCD
 with a superpotential, coupled to KK adjoint matter \cite{CaseroPT}.
The corresponding black hole is known only for the case $N_f=2N_c$.
 Otherwise stated we shall work in the Einstein frame and with $\alpha^\prime=1$.
The metric reads
\bea\label{metriczero}
ds^2_0 &=& e^{(\phi_0+r)/2} \Bigl[ d\vec x_4^2
+ N_c \Bigl( dr^2 + {1\over \xi} (d\theta^2+\sin^2\theta d\varphi^2)
+{1\over 4-\xi}  (d\bar\theta^2+\sin^2\bar\theta d\bar\varphi^2)\nonumber\\
&& +{1\over 4} (d\psi +\cos\theta d\varphi +  \cos\bar{\theta}
d\bar{\varphi})^2\Bigr) \Bigr], \eea for the ``zero temperature''
background and \bea\label{metrictemp} ds^2_T &=& e^{(\phi_0+r)/2}
\Bigl[ - {\cal F} dt^2 + d\vec x_{3}^2 + N_c
\Bigl( {1\over \cal F} dr^2 + {1\over \xi} (d\theta^2+\sin^2\theta d\varphi^2)\\
&&+{1\over 4-\xi}  (d\bar\theta^2+\sin^2\bar\theta d\bar\varphi^2)
 +{1\over 4} (d\psi +\cos\theta d\varphi +  \cos\bar{\theta} d\bar{\varphi})^2\Bigr) \Bigr]\,,
\qquad {\cal F} = 1- e^{(r_0-r)}\,,\nonumber \eea for the thermal one.
The angles run in $0\le \theta(\bar{\theta})  \le \pi,\ 0\le\varphi
(\bar{\varphi})\le 2\pi, \  0\le\psi\le 4 \pi$. The horizon of the
thermal solution is located at $r_0$; $t,\ x_i$ are dimensional
 coordinates, while the rest are dimensionless; $\xi$ is a free parameter that varies in the range $0< \xi < 4$.
The backgrounds include also a dilaton and a three form field which
are identical in both cases \footnote{Note that there is a typo in
equation (4.23) of \cite{CaseroPT}, see eq. (3.18) of the same
paper.} \bea
e^\phi&=&e^{\phi_0+r}\,,\\
F_{(3)}&=&-{N_c\over 4} \left( \sin\bar{\theta}  d\bar{\theta}
\wedge d\bar{\varphi} + \sin\theta d\theta \wedge d\varphi\right)
\wedge \left( d\psi + \cos\theta d\varphi + \cos\bar{\theta}
d\bar{\varphi} \right)\,. \label{threeform}\eea

These backgrounds are solutions of the Type IIB supergravity
equations of motion for $N_c$ color D5-branes wrapped on a
two-sphere, plus a DBI source to account for the backreaction of
$N_f$ D5 flavor branes. The latter are aligned along $r,\ \psi$ and
the Minkowski coordinates, and they are smeared in the other
directions. Being (\ref{metriczero}), (\ref{metrictemp}) (decoupled)
D5-brane solutions, they are asymptotic for large radius to
(compactified) Little String Theory backgrounds. As such, the black
hole shares with the thermal LST solution the main feature that its
temperature does not depend on $r_0$, \be\label{temperature}
T=\frac{1}{2 \pi \sqrt{N_c} }\,. \ee As a consequence, its free
energy is expected to vanish if the usual thermodynamic relations
are still valid for this system. This is indeed the case, as we will
explicitly compute in the following. Finally, the functional form of
the scalar correlator agrees with that presented in \cite{NarayanDR}
for LST.

\section{Thermodynamics and Holographic Renormalization}\label{termodyn}

In this section we will study the thermodynamics of the above model.
As already stressed, being the latter the only available critical model describing a finite temperature field theory with backreaction of the fundamental flavors, its thermodynamic analysis is surely worthwhile.

Moreover, it turns out to be a quite useful arena where to study the
holographic procedures in the presence of the flavor branes. In
fact, as usual when one evaluates thermodynamic quantities, such as
the free energy or the energy density, one deals with infinities.
There exist standard procedures in order to extract anyway the
relevant informations from supergravity, as measuring them with
respect to a reference background or by a subtraction mechanism
known as holographic renormalization. But in the case at hand the
system does not include only the supergravity action -- the latter
is indeed coupled to the (smeared) DBI source for the flavor branes.
Thus, it is not a-priori clear whether the known procedures for
supergravity can be applied to this case. We will instead show that
both methods give (the same) meaningful answers for the
thermodynamics of the system at hand.

Before coming to the backreacted ${\cal N}=1$ SQCD model in section \ref{sqcdtermo}, we will review some of the
underlying aspects one finds in near-extremal NS5-branes, where one does not deal with the flavor contribution.

\subsection{Preliminaries: Thermodynamics of LST}\label{seclst}

One of the reasons the LST model is interesting is the fact that it
is a non-gravitational theory believed to be dual to a certain
string theory background. In the decoupling limit of $N$ coincident
NS5-branes the string length $l_s$ is kept fixed while the string
coupling $g_s$ goes to zero. In that precise limit the theory
reduces to Little String Theory, or more precisely to $(2,0)$ LST
for type IIA NS5-branes and $(1,1)$ LST for type IIB NS5-branes
\cite{AharonyUB}. We are interested in the non-extremal case. The
decoupling limit is achieved by keeping $l_s$~fixed and taking
$g_s$~to zero while the energy above extremality is fixed. The
throat geometry described by this setting is \cite{CallanAT}
\be\label{lstgeo} ds^2 = e^{-\phi/2}\left[\left(1-{z_0^2\over
z^2}\right) dx_1^2+ \sum_{j=2}^6 dx_j^2+N \left( {dz^2\over
z^2-z_0^2 } +d\Omega_3^2\right) \right],\qquad e^{2 \phi}= {N\over
z^2}\,. \ee The extremal configuration is obtained by the limit
$z_0\to 0$. In the latter limit, (\ref{lstgeo}) represents a
five-brane whose world-volume is ${\bf R}^6$. The string propagation
in this geometry should correspond to an exact conformal field
theory \cite{KutasovUA} \be\label{cf}{\bf R}^{5,1} \times {\bf
R}_\phi\times SU(2)_N\,. \ee In addition to the previous fields
there is a NS-NS $H_{(3)}$ form along the $S^3$, $H_{(3)}= 2 N
\epsilon_3$. The geometry (\ref{lstgeo}) is regular as long as
$z_0\ne 0$ and (in order not to develop a conical singularity) the
period of the Euclidean time
 is chosen as
\be\label{temp} \beta= 2 \pi \sqrt{N}\,. \ee Notice that this value
is {\sl fixed} and independent of the black hole radius, and leads
to a complete degenerate thermodynamical phase space
\cite{MaldacenaYA}. It corresponds to the Hagedorn temperature of
superstring theory.

A basic thermodynamic quantity one would like to compute for the
thermal system corresponding to the background (\ref{lstgeo}) is the
free energy. It can be calculated as usual from the two terms in the
action
\be\label{I}
 {\cal I}={\cal I}_{\rm grav}
+{\cal I}_{\rm surf}\,. \ee The former is the Einstein-Hilbert
action \be\label{action} {\cal I}_{\rm grav}= {1\over  2
\kappa^2_{10} } \int_{\cal M} d^{10}x \sqrt{g} \left( R-{1\over 2}
\partial_\mu \phi \partial^\mu \phi  -{1\over 12} e^{-\phi}
 H_{(3)}^2\right)\,,
\ee while the latter is the surface contribution \cite{GibbonsUE}
\be\label{surf} {\cal I}_{\rm surf}= {1\over  \kappa^2_{10} }
\oint_\Sigma  K d\Sigma\,, \ee with ${\cal M}$ being a ten-volume
enclosed by a nine-boundary $\Sigma$. The surface contribution is
determined by the extrinsic curvature
 \be\label{K}
 K= \nabla_\mu n^\mu = {1\over\sqrt{g}} \partial_\mu \left( \sqrt{g}\, n^\mu \right)\,,
 \ee
 where $n^\mu$~is the boundary outward normal vector.
It is chosen at a constant cutoff value of the radial coordinate $r$ which
eventually will be taken to infinity, see for instance
\cite{Gubser:2001eg}.

To leading order, the free energy is just the classical action evaluated
on the solution of the equations of motion of the gravity-matter system, divided by the
inverse temperature $\beta$. In our case,
the action (\ref{I}) diverges as the radial cutoff is sent to infinity and hence needs
 to be renormalized.
In the following we perform it in two different ways and check the consistency of
the approaches.

\subsubsection{Reference background renormalization}

One way to renormalize the action (\ref{I}) is to subtract the corresponding quantity for
a reference background. This must be viewed as the equivalent of
 fixing the vacuum energy. The natural choice for the reference background
 is the extremal metric obtained from (\ref{lstgeo}) by sending $z_0\rightarrow 0$ and we shall proceed with it in the sequel.

We then evaluate expression (\ref{I}) on the background with a
radial cut-off $\R$, obtaining \be\label{io} {\cal I}_{e} =  {1\over
2 \kappa^2_{10}} \,  V_5 \,2 \pi^2 \int_0^{\beta^\prime}dt \left(-
\int_0^\R dz\, z\, +{9\over 2} \R^2 \right), \ee for the extremal
case and \be\label{itt} {\cal I}_{ne} =  {1\over  2 \kappa^2_{10}}
\,  V_5  \,2 \pi^2 \int_{0}^{\beta}dt \left(- \int_{z_0}^\R dz\, z
+{1\over 2} \left(9 \R^2 - 5 z_0^2\right) \right)\,, \ee for the
non-extremal one. In the expressions above, $V_5$ stands for the
volume of the ${\bf R}^5$
 along the
branes, the factor $2 \pi^2$ accounts for the volume of the ${\bf
S}^3$ and $\beta^\prime$ and $\beta$ are of course the periods of
the compactified Euclidean time directions.

Notice that both bulk radial integrands are the same (just $z$).
Thus, besides the surface contribution, differences arise from two
sources, both due to the integration region: {\sl i)} the radial
coordinate for the non-extremal case is subject to $z\ge z_0$ while
in the extremal case $z\ge 0$. {\sl ii)} The periodicity $\beta$ of
the thermal circle in the non-extremal case is fixed to $\beta=2 \pi
\sqrt{N}$; in the extremal solution $\beta^\prime$ is not fixed in
the solution, but in order to make sense of the subtraction one must
carefully adjust it in order that the lengths of the two thermal
cycles at the cutoff coincide. This is done by setting $\beta^\prime
= \beta \sqrt{\vert g_{00}\vert }\,,$ where $g_{00}$ is evaluated at
the cutoff.

The result is finally obtained by sending the cutoff to infinity
\cite{BerkoozMZ} \be {\cal I} =  \lim_{\R\rightarrow
\infty}\bigl[{\cal I}_{ne} - {\cal I}_{e}\bigr]= \lim_{\R\rightarrow
\infty} \frac{(2\pi)^3}{2 \kappa^2_{10}}\sqrt{N}\,  V_5 \left[(2
\R^2-z_0^2)-2 \R \sqrt{\R^2-z_0^2}\right]=0\,. \ee The free energy of
the system vanishes.

\subsubsection{Holographic renormalization}

While the procedure just explained is successful, one is still
restricted to use a reference background and may wonder whether the
result depends on its choice. Moreover in some cases this reference
background may not exist or just, as in the previous example,
contains a singularity.

A second procedure overcomes these possible objections by using an
effective approach \cite{SkenderisUY,KarchMS}. One makes use of the
fact that the radial coordinate transverse to the branes is related
to the energy scale in the dual field theory \cite{PeetWN}. For
example, the string propagating on the geometry (\ref{cf}) is
related to a dual field theory, in the sense that gravity quantities
at a given radius $\R$ correspond to field theory observables at a
fixed energy (related to $\R$). The field theory observables, that
live on the world volume of the NS5 branes, eventually must be
rendered finite by renormalizing them. In order to do so in the
gravity dual, one identifies the functional form of all possible
sources of divergences that can be generically obtained on the world
volume of the NS5. Writing down the metric (\ref{lstgeo}) in the
form \be\label{lstuv} ds^2=  ds^2_{6+1} + e^{2 \sigma(z)} L^2\,
d\Omega_3^2\, , \ee the prescription for the generic counterterm is
\cite{MateosVN} \be\label{ct} {\cal I}_{ct}= \frac{A\, L^2\,
\Omega_3}{\kappa_{10}}   \int_{\tilde{{\cal M}}} d^6 x \,
\sqrt{\tilde{g}_6} \, e^{3 \sigma(z)} \, e^{B \phi(z)}\, , \ee where
$\tilde{g}_6$ denotes the metric \footnote{Note that, having an
undetermined power of the dilaton, there is no difference between
the Einstein and the string frame.} induced on the brane world
volume $\tilde{{\cal M}}$, computed at the cutoff $\R$. The term
$e^{3 \sigma(z)}$ comes from the coefficient of the three-sphere, of
volume $\Omega_3$, in the
 determinant of the full metric.
The constants $A, B$ are then tuned in order to cancel the
divergences in the action ${\cal I}_{ne}$ (\ref{itt}). In our case
the choice $ A=2 ,\,
 B=-{1\over 4}\,
$ is uniquely selected by the requirement of absence of divergences
in the $\R \rightarrow \infty$ limit, and leads to the same result
as before
 \be\label{diffbis}
{\cal I} =  \lim_{\R\rightarrow \infty}({\cal I}_{ne} - {\cal
I}_{ct}) = 0\,. \ee

\subsubsection{Energy and Entropy}

Since the free energy of the system, $F={\cal I}/\beta$, is vanishing, from the usual thermodynamic relation $F=E-T S$ one expects that the energy is proportional to the entropy.

For a stationary spacetime admitting foliations by spacelike
hypersurfaces $\Sigma_t$ the conserved energy enclosed in a shell is
given by the ADM relation \cite{ArnowittHI,HawkingFD} (the Newton
constant $16 \pi G_N=2 \kappa_{10}^2$ is explicitated for latter
convenience)
 \be\label{adm} E=  - {1\over 8 \pi G_{10}} \oint_{\Sigma_t}
(K-K_0) \sqrt{\vert g_{00}\vert }\, d \Sigma_t\,, \ee
 where now the non-extremal $K$ (extremal $K_0$) extrinsic curvature is obtained at a
fixed time slice. One can compute the energy density for the LST
background (\ref{lstgeo}) by using the definition (\ref{K}),
obtaining \be\label{denen}e \equiv {E\over V_5} =  {\pi\, z_0^2
\over 4\, G_{10}}\,. \ee Furthermore, the entropy density calculated
from the area of the black hole horizon with the Bekenstein-Hawking
relation \be S={{\rm Area}\over 4 G_{10}} \ee gives
\be\label{enden}s\equiv {S\over V_5} = {\pi^2 \sqrt{N}\,z_0^2 \over
2\, G_{10}}\,. \ee Thus, using the fact that $T=\frac{1}{2 \pi
\sqrt{N}}$, it follows that the LST background satisfies the usual
thermodynamic relations and $E=TS$.

\subsection{Thermodynamics of ${\cal N}=1$ SQCD-like theories}\label{sqcdtermo}

In the previous section we have reviewed how the reference background and holographic renormalization work in the simple case of LST, where one does not have any flavor branes.
We now turn to the more involved case of the flavor-backreacted metric (\ref{metrictemp}), dual to SQCD-like theories with $N_f=2N_c$ flavors.
In order to study its thermodynamics, we follow the same steps as in the LST model, but including now the DBI term for the flavor branes.
We shall find a completely consistent picture by using the straightforward extensions of the usual procedures.

As a first thing, we compute the free energy from the Euclidean
action. The latter reads in this case \be\label{action2} {\cal I} =
{\cal I}_{\rm grav}  + {\cal I}_{\rm flavor} + {\cal I}_{\rm
surf}\,, \ee where the first and third terms are the standard ones,
\be\label{grac} {\cal I}_{\rm grav} = {1\over 2 \kappa^2_{10}}
\int_{\cal M} d^{10}x \sqrt{-g} \left( R-{1\over 2} \partial_\mu
\phi \partial^\mu\phi -{1\over 12} e^{\phi} F_{(3)}^2\right), \qquad
{\cal I}_{\rm surf}= {1\over  \kappa^2_{10} } \oint_\Sigma  K
d\Sigma\,, \ee where $K={1\over\sqrt{g}} \partial_\mu \left(
\sqrt{g}\, n^\mu \right)$. ${\cal I}_{\rm flavor}$ is instead given
by the inclusion of fundamental matter \cite{CaseroPT} through a
family of $N_f$ D5-branes originally extended along the
$t,x_1,x_2,x_3,r,\psi$ directions and finally evenly smeared in the
extra compact directions $\theta, \varphi, \bar\theta, \bar\varphi$.
It is given by \be\label{flavor}{\cal I}_{\rm flavor}= -{T_5
N_f\over (4\pi)^2} \int_{\cal M} d^{10}x \sin\theta \sin\bar{\theta}
e^{\phi/2} \sqrt{-\tilde{g}_6}+ {T_5 N_f\over (4\pi)^2}\int_{\cal M}
{\rm Vol}({\cal Y}_4) \wedge C_{(6)}\,, \ee where in string units
$\displaystyle T_5={1/ (2 \pi)^5}$.

\subsubsection{Free energy from the reference background renormalization}

As in the LST model the temperature of our background (\ref{metrictemp}) is fixed to $T=\frac{1}{2 \pi \sqrt{N_c} }$ and so it does not depend
on the mass, or size,
of the black hole. We expect in such a case, as we know from the study of the LST, an energy
proportional to the entropy and therefore a vanishing free energy. We shall show that this is
indeed the case, by using as reference background the zero temperature one (\ref{metriczero}) in order to renormalize the infinite free energy of the thermal background (\ref{metrictemp}).

Let us first consider the ``bulk'' contributions to the action
${\cal I}$ \be {\cal I}_{\rm bulk}\equiv {\cal I}_{\rm grav}+ {\cal
I}_{\rm flavor}\,. \ee As a first thing, one notices that the
specific structure of the three-form (\ref{threeform}) guarantees
that the Chern-Simons term in (\ref{flavor}) vanishes (both in the
zero temperature and in the finite temperature cases). It is then a
matter of straightforward computation to evaluate the remaining
parts of ${\cal I}_{\rm bulk}$ on the background. In the case of the
finite temperature solution (\ref{metrictemp}) one gets, for the
integrand in ${\cal I}_{\rm bulk}$, \be\label{integrandT} {\rm
Integrand}_{\rm bulk} = {{e^{2(r+\phi_0)}N_c^2
\sin\theta\sin\bar\theta}\over{(4-\xi)\xi}}\,, \ee which is
independent of $r_0$. As a consequence, the same integrand will be
obtained for the zero temperature solution (\ref{metriczero}).  This
is a very typical situation that we have already seen in the LST and
it also happens in the case of the AdS black hole \cite{WittenZW}.
This means that the effects of the background subtraction, as
regards the bulk contributions, are only due to differences in the
integration region. There are two such differences. In the finite
temperature case the radial integration goes from $r_0$ to the
cutoff $\R$ whereas in the zero temperature case it starts at $r=
-\infty$. The second difference is the range of integration of the
Euclidean time variable. In the zero temperature case this range,
$\beta^\prime$, is arbitrary and must be adjusted to the value
$\beta^\prime= \beta \sqrt{1-e^{2(r_0-\R)}}$, where $\beta=2 \pi
\sqrt{N_c}$, such that the geometry of this background coincides
with that of the finite temperature one at the cutoff $\R$. With
these ingredients one can compute the finite temperature bulk
contribution, subtract from it the zero temperature contribution at
the cutoff $\R$ and finally send $\R$ to infinity, getting
 \be\label{bulk}{{\cal I}_{\rm bulk}\over V_3}= {{(2\pi)^4 e^{2(r+\phi_0)}N_c^{5\over 2} }\over
{2\kappa_{10}^2(4-\xi)\xi}}\,, \ee with $V_3= {\rm Vol}({\bf R}^3)$.

Next we move to the boundary contribution ${\cal I}_{surf}$. Using
again $\beta^\prime$ and $\beta$ for the thermal cycles and
performing the subtraction of the contributions from the two
backgrounds we get, after sending $\R$ to infinity,
\be\label{bound}{{\cal I}_{\rm surf}\over V_3}=
-{{(2\pi)^4e^{2(r+\phi_0)}N_c^{5\over 2} }\over
{2\kappa_{10}^2(4-\xi)\xi}}\,. \ee Thus the free energy density
vanishes \be\label{fe} f= {1\over \beta V_3}({\cal I}_{\rm bulk}
+{\cal I}_{\rm  surf})=0\,, \ee as expected. Of course, the novel
ingredient, that is the DBI term for the flavors ${\cal I}_{\rm
flavor}$, is crucial in order to obtain the exact coefficient in
(\ref{bulk}) such that the free energy is zero.

In a sense, even if the result is the expected one, our computation
might look somewhat suspicions, or problematic, for the zero
temperature background has a singularity at $r\rightarrow -\infty$.
Although being classified as a ``good" singularity, it signals the
breakdown of the supergravity approximation, and in view of that one
may wonder about the legitimacy of the result. So, in order to test
the solidity of our result we shall do two independent checks. On
one hand, we can compute separately the energy and the entropy
associated with our finite temperature background and check whether
the free energy vanishes. This will be done later. On the other hand
we can use a holographic counterterm subtraction instead of the
background subtraction. We devote the next subsection to this last
issue.

\subsubsection{Free energy from holographic renormalization}

We are now going to renormalize the action (\ref{action2}) using the
holographic renormalization prescription. While for the
gravitational part this is fairly standard and we have reviewed it
in the previous case of LST, for the flavor part its implementation
is still not a fully tested procedure. Namely, while it has been
studied for the renormalization of brane actions on curved
backgrounds, as far as we know it has never been used for the case
where the brane action is a source of the gravitational background.
Thus, here we provide the first full-fledge example of such
procedure, where we confirm that the prescriptions in
\cite{KarchMS,SkenderisUY,MateosVN} give consistent and unique
results.

As in the previous section, it is a matter of straightforward
computation to plug the finite temperature solution
(\ref{metrictemp}) in the action (\ref{action2}) and integrate up to
a radial cutoff $\R$. The result for the usual gravitational part
(\ref{grac}), ${\cal I}_{\rm gravity}\equiv {\cal I}_{\rm
grav}+{\cal I}_{\rm surf}$, reads \be\label{freeTgrav}{\cal I}_{\rm
gravity}={{ e^{2 \phi_0} N_c^{5\over 2}V_3}\over {\pi^3 }} \cdot
\frac{e^{2 \R}[16+(4-\xi)\xi]-e^{2
r_0}[8+(4-\xi)\xi]}{8(4-\xi)\xi}\,, \ee while the contribution of
the flavor branes (\ref{flavor}) is \be\label{freeTflav}{\cal
I}_{\rm flavor}= {{e^{2 \phi_0} N_c^{5\over 2}V_3}\over {\pi^3
}}\cdot \frac{(e^{2 r_0}-e^{2 \R})}{8}\,. \ee Each of the two pieces
must be renormalized.

For what concerns the gravitational part ${\cal I}_{\rm gravity}$
(\ref{freeTgrav}), as we saw in the LST case the general holographic
renormalization setup concerns a background generated by a stack of
Dp-branes with a sphere transverse to the branes, which can be
dimensionally reduced (see Appendix E of \cite{MateosVN}). There are
some notable differences with our present case. In our D5-brane case
the transverse space has a more complicated structure, a fibered
$S^3$, and, in addition, the D5-branes wrap a two-cycle. These
differences notwithstanding, we just take a slight generalization of
the proposal in \cite{MateosVN} in the sense that there should exist
an effective counterterm which, when written before undertaking any
dimensional reduction of the transverse space, must be of the form
\be \label{mmt} {\cal I}_{\rm ctgravity}= {A\over L \kappa_{10}^2}
\int_{\partial {\cal M}}  d^{9}x \sqrt{h}\, e^{B\phi}\,, \ee where
$L=\sqrt{N_c}$ is the scale factor in front of the compact part of
the metric (see (\ref{lstuv}) for comparison), $\partial {\cal M}$
is the boundary of the manifold at $r=\R$, $h$ is the boundary
metric, and $A$ encompasses a global constant which, together with
the constant $B$, is determined in order to cancel the divergences
in (\ref{freeTgrav}). Cancelation of the large $\R$ divergence in
${\cal I}_{\rm gravity}+{\cal I}_{\rm ctgravity}$ then fixes
$A=-[16+(4-\xi)\xi]/8$ and $B=-1/4$. Thus \be {\cal I}_{\rm
gren}\equiv \lim_{\R \rightarrow \infty}({\cal I}_{\rm
gravity}+{\cal I}_{\rm ctgravity}) =
-\frac{e^{2(r_0+\phi_0)}N_c^{5/2}V_3}{16 \pi^3}\,. \ee

Analogously, for the flavor part ${\cal I}_{\rm flavor}$
(\ref{freeTflav}) we use the prescription in \cite{MateosVN}, taking
the counterterm \be \label{mmt2}{\cal I}_{\rm ctflavor}= 2\pi T_5
N_f L^2 A' \int  d^{4}x \sqrt{\gamma} e^{2\sigma'}e^{B'\phi}\,, \ee
where $\gamma$ is the induced metric on the 4d part of the
world-volume, $2\pi e^{2\sigma'}=2\pi e^{(r+\phi_0)/4}/2$ comes from
the wrapped $\psi$-direction, and $A', B'$ must be adjusted in order
to cancel the divergence in ${\cal I}_{\rm flavor}+{\cal I}_{\rm
ctflavor}$. The latter requirement then fixes $A'=-1$ and $B'=3/4$.
The renormalized flavor contribution is thus \be {\cal I}_{\rm
fren}\equiv \lim_{\R \rightarrow \infty}({\cal I}_{\rm flavor}+{\cal
I}_{\rm ctflavor}) = \frac{e^{2(r_0+\phi_0)}N_c^{5/2}V_3}{16 \pi^3}\,,
\ee and so the free energy density is vanishing
\be\label{renormfreeE} f = {1\over\beta\, V_3}({\cal I}_{\rm
gren}+{\cal I}_{\rm fren}) = 0\,. \ee This agrees with the result
previously obtained by using the background subtraction method, thus
putting the whole procedure on a solid ground.

Let us note that we could have joined the two separate pieces, the
standard gravitational one and flavor one, in a unique term, for
which we could have used just one counterterm. The latter would have
of course produced again a vanishing free energy. Nevertheless,
since this procedure would have hidden the structure of the
counterterms, above we choose to present the separate
renormalization of the two pieces.

\subsubsection{Energy and entropy}

A direct procedure to check the vanishing of the free energy is to
compute independently the energy and the entropy. We use the
relation (\ref{adm}) to compute the conserved ADM energy. The energy
density turns out to be \be\label{en} e= {1\over 8\pi
G_N}{e^{2(r_0+\phi_0)}N_c^2 (4\pi)^3\over
2(4-\xi)\xi}=\frac{8\lambda^4}{(4-\xi)\xi}T^4\,, \ee where we defined
$\lambda\equiv e^{r_0+\phi_0}N_c$, which is the quantity that must
be fixed and large in order for the gravitational description of the
system (\ref{metrictemp}) to be reliable. In addition we used the
fact that in string units $\alpha^\prime=1$ and in the Einstein frame the
Newton constant is given by
$G_N=\kappa_{10}^2/8\pi=2^3\pi^6/e^{2(r_0+\phi_0)}$.

As for the entropy density, as a quarter of the area of the horizon
of the metric (\ref{metrictemp}) at $r=r_0$, it is just \cite{noi}
\be\label{entr}s= {1\over 4 G_N}{e^{2(r_0+\phi_0)}N_c^{5\over 2}
(4\pi)^3\over 2(4-\xi)\xi}=\frac{8\lambda^4}{(4-\xi)\xi}T^3\,. \ee
Thus the free energy density $f=e-Ts$ vanishes, as expected.

Note the nice feature that the energy and entropy density are
invariant under the change $\xi \rightarrow 4-\xi$, which is
supposed to correspond to a Seiberg duality for the field theory
dual to the zero energy solution \cite{CaseroPT}.\footnote{We thank
C. Nunez for this comment.} This is of course a trivial consequence
of the invariance of the background under this transformation
(together with $(\theta, \phi)\leftrightarrow (\bar\theta,
\bar\phi)$).

All in all, the thermodynamics of this model is very similar to the
LST one, the temperature is fixed irrespectively of the horizon size
and the free energy vanishes identically. The thermal system is
described by a black hole at the fixed Hagedorn temperature. In LST
it was argued, by studying the first string corrections to this
situation or by studying compactified systems, that there are
thermal instabilities, as the specific heat turns out to be negative
\cite{KutasovJP,Rangamani:2001ir,BuchelDG,Gubser:2001eg,Bertoldi:2002ks}.
Given the similarities of the two models, one can expect similar
instabilities in our thermal background (\ref{metrictemp}) too, but
a much more involved analysis, which is outside the scope of this
paper, would be required in order to check this statement.

\section{Jet quenching parameter}\label{quenching}

The solution (\ref{metrictemp}) we are considering in this note is
the only known ten dimensional black hole dual to a 4d finite
temperature field theory that includes the backreaction of many
flavor degrees of freedom. This is a sufficient motivation for the
investigation of its properties, despite its thermodynamics
resembles the Little String Theory one and may eventually reveal
instabilities. In fact, as stressed in the Introduction, there has
been evidence that string theory may provide some insight into the
study of properties of the Quark Gluon Plasma. Here we are
interested in the effects of the backreaction of fundamental flavors
on one specific plasma observable, namely the jet quenching
parameter.

In \cite{noi} the authors calculated for the background
(\ref{metrictemp}) some quantities relevant to the explanation of
the huge energy loss of colored probes in the QGP, namely the jet
quenching parameter \cite{LiuUG} and the relaxation time
\cite{HerzogGH,GubserBZ}. While the result for the latter turns out
to be quite similar to the ones in previous ``unflavored'' string
models, the jet quenching parameter is unexpectedly vanishing. This
is not an effect of the presence of fundamental flavors, since it is
a common features to all the theories coming from D5-branes
\cite{noi}. But, the relevant observation for our scope is that in
order to perform the computation, a specific string configuration
was used, introduced in \cite{LiuUG} for the $AdS_5 \times S^5$
black hole dual to ${\cal N}=4$ SYM. The corresponding string
extends from infinity up to the horizon of the black hole. In
\cite{argy} it was observed that apparently that configuration is
not the minimal energy one. As such, the path integral is dominated
by other configurations and accordingly the jet quenching parameter
should be calculated using the latter.

Thus, considering this possibility, in this section we perform a
careful analysis of the string configurations that could be used to
calculate the jet quenching parameter, as proposed in \cite{argy}.
We will consider the set-up in which we start from strings at
velocity $V\neq 1$ and finite mass $m$ of the dual quarks, with
space-like world-sheet, and take the limit of $V\to 1$ and infinite
$m$ at the end, thus giving to the final result a physical origin.
As we will see, though, among all the possibilities, the only
configuration that seems to give a sensible physical result is the
one used in \cite{noi}, that gives zero jet quenching parameter.

To be concrete, the jet quenching parameter is calculated from
string theory as the coefficient of the $L^- L^2$ term in the action
for a macroscopic string, spanning a Wilson loop with a light-like
dimension $L^-$ much larger that the spatial one, of size $L$
\cite{LiuUG}. It corresponds to the infinite mass, infinite velocity
limit of a string describing a quark-antiquark pair. The latter
provides a good measure of the jet quenching parameter when it has a
space-like world-sheet \cite{liu3}.

Thus, consider a test string, in the static gauge, representing a dipole moving with velocity
$v$ along the $x_1$ direction, perpendicularly to its extension along $x_2$.
We will work in the string frame and with the radial coordinate $z= e^{r\over 2}$, such that the metric (\ref{metrictemp})
reads
\bea\label{metrictempz}
ds^2_T &=& e^{\phi_0}z^2 \Bigl[ - {\cal F} dt^2 + d\vec x_{3}^2 + N_c
\Bigl( {4\over z^2\cal F} dz^2 + {1\over \xi} (d\theta^2+\sin^2\theta d\varphi^2)\\
&&+{1\over 4-\xi}  (d\bar\theta^2+\sin^2\bar\theta d\bar\varphi^2)
 +{1\over 4} (d\psi +\cos\theta d\varphi +  \cos\bar{\theta} d\bar{\varphi})^2\Bigr) \Bigr],
\qquad {\cal F} = 1- \frac{z_0^4}{z^4}\,,\nonumber \eea and of course
$z_0$ is the position of the horizon. The relevant string
configuration is then given by \be t=\tau, \qquad x_1=v \tau,\qquad
x_2=\sigma, \qquad z(\sigma)\,. \ee The string is attached to a probe
``flavor'' D5-brane, whose radial position $z_b$ is dual to the
quark mass $m$. The separation of the quark-antiquark pair is
denoted by $L$. The only stable configuration is that which keeps
the plane where the string lies perpendicular to the velocity.

As pointed out in \cite{argy}, and can be deduced from
(\ref{metrictemp}), $v$ is not the proper dipole velocity. The true
velocity of the dipole is $V= {v/\sqrt{1-{z_0^4 / z_b^4}} }$. We are
interested in spacelike configurations on which the lightlike limit
$V\to 1$ will be taken. This means that $v\to 1$ and $ z_b\to
\infty$.\footnote{For $v<1$ there is in principle the finite case
when $z_b^4 = {z_0^4\over 1-v^2}$, but this case will be discarded
in due time. } It will make a difference whether the limit for $v$
is taken form below $v\to 1^-$ of from above $v\to 1^+$. The
spacelike condition $G_{\tau\tau} \equiv g_{\mu\nu}\partial_\tau
x^\mu \partial_\tau x^\nu >0$ will make the induced action on the
world-sheet, $S= {1\over2\pi} \int d\tau d\sigma \sqrt{-G}$,
imaginary. Therefore $e^{i\,S}$ will be a real exponential.
Following \cite{argy} we will take the minus sign for the exponent
$e^{i\,S}=e^{-S_r}$. Henceforth we will work directly with $S_r$.

\subsection{Spacelike configurations with $v\to 1^-$}


The induced action on the world-sheet is
\be\label{indm} S_r= {1\over2\pi} \int d\tau d\sigma
\sqrt{G}={1\over2\pi} \int d\tau d\sigma {e^{\phi_0}\over
\gamma}\sqrt{(z_0^4\gamma^2 -z^4) \left(1 + 4N_c  z^2 z^{\prime 2}
{1\over z^4-z_0^4}\right)}\,, \ee where we have introduced the
standard notation $\gamma^2=1/(1-v^2)$.

Let us first analyze the possible location of the turning point for
the string configuration. The $\sigma$-independent conserved
quantity of (\ref{indm}) is \be {z_0^4 \gamma^4 -z^4\over
{4\,N_c\,z^2\,z^{\prime 2}\over z^4-z_0^4}+1} \equiv h > 0\,. \ee
Thus, \be z^{\prime 2} = \Big({z_0^4\gamma^2-z^4-h \over
h}\Big)\Big({z^4-z_0^4\over 4\,N_c\,z^2}\Big)\,. \ee Since $z^4\geq
z_0^4$ for the string configuration, obviously one needs $z^4\leq
z_0^4\gamma^2-h$ to guarantee $z^{\prime 2}\geq 0$. Defining $z_t$
by $z_t^4\equiv  z_0^4\gamma^2-h\geq z^4\geq z_0^4$, one has \be
z^{\prime 2} = \Big({z_t^4-z^4 \over
z_0^4\gamma^2-z_t^4}\Big)\Big({z^4-z_0^4\over 4\,N_c\,z^2}\Big)\,, \ee
with $z_0\leq z\leq z_b\leq z_t \leq z_0\sqrt{\gamma}$, where $z_b$
is the location of the brane.

The possible turning points, for which $z^{\prime 2} = 0$, can only be $z_0$ or $z_t$.
Let us examine both cases.

 \medskip

\subsubsection{The {\sl down} configuration}

First consider the case when the string configuration reaches the
horizon at $z_0$, which will be the turning point. Using variables
$z= y z_0$, the distance between the two quarks at radial coordinate
$z_b$ is \be\label{length} {L\over \beta} ={1\over \beta}\int
d\sigma= {2\over \beta}\int_1^{y_b} {d y\over y'}= {2\over
\pi}\alpha \gamma \int_1^{y_b}d y {y\over\sqrt{(y_t^4-y^4)(y^4-1)}}\,,
\ee where $\beta$ is the inverse temperature and the definition
$\alpha^2= 1 - { y_t^4 / \gamma^2}\leq 1$ has been used. The action
can be written as \be\label{realaction} S_r= {1\over 2\pi}
{e^{\phi_0}\over \gamma} 4 \sqrt{N_c}\,\hat t\, z_0^2 \int_1^{y_b} d
y {y(\gamma^2-y^4)\over\sqrt{(y_t^4-y^4)(y^4-1)}}\,, \ee where
$\hat{t}$ is the time interval the dipole propagates.

In the limit $\gamma\to \infty$, while keeping $y_b$ finite, one
must also send $y_t\to \infty$ in order to keep $L$ finite. In this
limit, $L$ becomes \be {L\over \beta} \approx {2\over
\pi}{\alpha\over \sqrt{1-\alpha^2}} \int_1^{y_b}d y
{y\over\sqrt{y^4-1}} = {\alpha\over \pi \sqrt{1-\alpha^2}} \,{\rm
arccosh}\,(y_b^2)\,. \ee Notice that in a next step we will send
$y_b\to \infty$ and then $\alpha$ must vanish in order to keep $L$
finite. In the same limit, $S_r$ becomes \be S_r \approx {1\over
2\pi}\, {e^{\phi_0}\over \sqrt{1-\alpha^2}} 4 \sqrt{N_c}\,\hat t\,
z_0^2 \int_1^{y_b}d y {y\over\sqrt{y^4-1}} = {2\pi\lambda\,\hat
t\,\over \beta^2}{L\over \alpha}\,, \ee where $\lambda\equiv
e^{\Phi_0}\,z_0^2\, N_c $. To this action we must subtract the
infinite mass quark contribution associated with the action $S_0$
for two straight strings stretching from the brane to the horizon
\be\label{sup0}
 S_0= {1\over 2\pi
}{e^{\phi_0}\over \gamma} 4 \sqrt{N_c}\,\hat t\, z_0^2 \int_1^{y_b}d
y\,y\,\sqrt{\gamma^2-y^4\over y^4-1} \approx {2\pi\lambda\,\hat
t\,\over \beta^2}\,\sqrt{1-\alpha^2 }\,{L\over \alpha}\,. \ee In the
required limit $\alpha\to 0$, the subtraction gives a vanishing
renormalized action \be S_{ren}=\lim_{\alpha\to 0}(S_r-S_0) = 0\,.
\ee This result also holds if both limits, $\gamma\to\infty$ and
$y_b\to\infty$ are taken simultaneously -- with
$y_b\leq\sqrt{\gamma}$. Thus, being the coefficient of this
renormalized action identified in \cite{LiuUG} with the jet
quenching parameter $\hat q$, the conclusion is that for this
configuration $\hat q=0$, in accordance with what argued in
\cite{noi}. The main concern is if this result can be altered by
possible different configurations.

\medskip

\subsubsection{The {\sl up} configuration}

Now consider the turning point at $y_t \geq y_b$. With  the
approximation $y_b\gg 1$, which is always legitimate because at the
end we send $y_b \to \infty$, the distance between the quarks is
\be\label{dipo} {L\over \beta} \approx {2 \over \pi}  \alpha \gamma
\int^{y_t}_{y_b} {dy\over y \sqrt{(y_t^4-y^4)}}\,, \ee and the action
of the configuration becomes \be\label{acd} S_r \approx
{4\,\lambda\,\hat{t}  \over \beta \gamma} \int^{y_t}_{y_b} {
(\gamma^2-y^4)\ dy\over y\ \sqrt{(y_t^4-y^4)}}\,. \ee Now we have
$$1<y_b\leq y \leq y_t \leq \sqrt{\gamma}.$$

Let us define $\eta = {y_t\over y_b} \geq 1$. Then the parameters
$(y_b,y_t,\gamma)$ can be traded for $(y_b,\alpha,\eta)$. The
distance $L$ becomes \be {L\over \beta} \approx {\alpha  \over \pi}
{1 \over \sqrt{1-\alpha^2}}\,{\rm arccosh}(\eta^2)\,, \ee out of which
we determine \be\label{etaeq} \eta(L,\alpha) =
\Big(\cosh({\pi\,\sqrt{1-\alpha^2}\,L \over  \beta
\alpha})\Big)^{1\over 2}\,. \ee $S_r$ is then written as
\begin{eqnarray}
\label{Sup}
S_r &\approx&   {2\lambda\,\hat{t} \over \beta} \,
\Big(-\sqrt{1-\alpha^2}\sqrt{1-{1 \over \eta^4}} +
{{\rm arccosh}(\eta^2)\over \sqrt{1-\alpha^2}}\Big)
\\ \nonumber
&=& {2\lambda\,\hat{t} \over \beta} \Big(-\sqrt{1-\alpha^2}
\tanh({\pi\,\sqrt{1-\alpha^2}\,L \over  \beta \alpha}) + { \pi \over
\beta}{L\over \alpha}  \Big).
\end{eqnarray}

Notice, from the definitions of $\eta$ and $\alpha$, the relation
\be\label{quot} {y_b^4 \over\gamma^2}= {1-\alpha^2 \over
\eta^4}={1-\alpha^2\over \Big(\cosh({\pi\,\sqrt{1-\alpha^2}\,L \over
\beta \alpha})\Big)^2}\,, \ee that will prove useful in the following.

Our natural free parameters are $\gamma, L, y_b$, with the
restriction $y_b^2\leq \gamma$. The values of $\alpha$ and $\eta$
are then read off from (\ref{quot}) and (\ref{etaeq}). Let us
examine the possible limits leading to $V\to 1$. One can easily
discard the case with $v<1$ and finite $y_b$. This setting implies
$y_b^2=\gamma$, but, according to (\ref{quot}), this is to set
$1-\alpha^2= \Big(\cosh({\pi\,\sqrt{1-\alpha^2}\,L \over  \beta
\alpha})\Big)^2$ which is clearly incompatible.

Our task now is to take both limits $v\to 1^-$ (that is,
$\gamma\to\infty$) and $y_b\to\infty$ subject to the restriction
$y_b \leq \sqrt{\gamma}$.

\begin{itemize}
\item[{\it{(a)}}] Let us then first take the limit $\gamma\to \infty$.
Inspection of (\ref{quot}) shows that there are two ways to do it,
while keeping $y_b$ finite: either by sending
$\alpha\to 1$, which implies $\eta\to 1$, or by sending
$\alpha\to 0$, which implies $\eta\to \infty$. Let us examine more closely both cases, always keeping $y_b$ finite.
\begin{itemize}
\item[{\it{(a.1)}}]
If $\alpha\to 1$, the consequence  $\eta\to 1$ dictates $y_t\to
y_b$, that is, the string configuration does not actually leave the
brane. The action written above remains finite in this limit, $S_r=
{2\pi\lambda\hat{t} \over \beta^2}\, L $. This result is independent
of the value $y_b$. The standard procedure of subtracting from $S_r$
the action $S_0$ corresponding to two straight lines  corresponding
to the two bare quark masses from $y_b$ to the horizon, will yield a
negative infinite value for the renormalized action when $y_b$ is
sent to infinity, since $S_0 \sim {\rm arccosh}(y_b^2)$.

\item[{\it{(a.2)}}]
If we take the opposite limit, $\alpha\to 0$, then $\eta\to \infty$
and therefore the turning point goes to infinity, $y_t\to \infty$.
Now the action is infinite and some subtraction is mandatory to
obtain a finite result, but again it seems not to exists any finite
subtraction because the infinite piece in $S_r$ goes as $L\over
\alpha$ whereas the action for the two straight lines goes as
$\ln(y_b)$.\footnote{Trying to adjust both limits, $\alpha\to 0$ and
$y_b\to \infty$, in order to obtain a finite renormalized action is
an artificial procedure which lacks of any reasonable support.}
Since we have to take first the $\alpha\rightarrow 0$ limit at fixed
$y_b$, we find a positive infinite result in this case.
\end{itemize}

\item[{\it{(b)}}]
The remaining case to explore is to take both limits, $\gamma\to
\infty$, $y_b\to \infty$, simultaneously such that (\ref{quot}) does
not go to zero. So we fix $\alpha$ with $0<\alpha<1$. In such case
we run into the same problem we met before. Now $S_r$ is finite
before subtraction, but the subtraction piece is an infinite
quantity when $y_b\to \infty$, and so we end up again with a
(negative) infinite value for the renormalized action.
\end{itemize}

\bigskip

In conclusion, none of the ``{\sl up}'' configurations with $v<1$
allows for a lightlike limit with a finite renormalized action. The
configurations with a negative infinite value for the action would
dominate the path integral. Nevertheless, since the subtraction
piece just gets rid of the infinite mass of the quarks in the limit
$y_b\to \infty$, the fact that, for these configurations, $S_r$ is
finite before subtraction means that the infinite quark masses are
already canceled by an infinite binding energy. Such configurations,
which have infinite action, have hardly a physical meaning in the
dual field theory, so we discard this possibility.


\subsection{ Spacelike configuration with $v\to 1^+$}

We introduce the parameter $\tilde \gamma^2 = {1 \over v^2-1}$. Let
us first analyze the possible location of the turning point for the
string configuration. The $\sigma$-independent conserved quantity of
the action is \be {z_0^4\tilde  \gamma^4 + z^4\over
{4\,N_c\,z^2(z^\prime)^2\over z^4-z_0^4}+1} \equiv h > 0\,. \ee Thus,
\be (z^\prime)^2 = \Big({z^4+ z_0^4\tilde \gamma^2-h \over
h}\Big)\Big({z^4-z_0^4\over 4\,N_c\,z^2}\Big)\,. \ee If $h<z_0^4+
z_0^4\tilde \gamma^2$, the only possible turning point is at $z_0$,
because  $z\geq z_0$. The critical case $h=z_0^4+ z_0^4\tilde
\gamma^2$ makes $z_0$ to be a simple zero of $z^\prime$; this has
the consequence that the distance between the two quarks can not be
kept finite, and thus we discard this configuration. Finally, for
$h>z_0^4+ z_0^4\tilde \gamma^2$ one can define $z_t^4\equiv
h-z_0^4\tilde \gamma^2 > z_0^4$, and we can write in accordance \be
(z^\prime)^2 = \Big({z^4-z_t^4 \over z_0^4\tilde
\gamma^2+z_t^4}\Big) \Big({z^4-z_0^4\over 4\,N_c\,z^2}\Big)\,, \ee
with the only turning point in $z_t$ because if it were $z_0$, there
will always exist at least some region where $(z^\prime)^2$ becomes
negative, no matter where is the location $z_b$ of the brane (remind
that $z_0 < z_t\,$).

\medskip

\subsubsection{The {\sl down} configuration}

Let us first consider
the case when the string configuration reaches the horizon at $z_0$, which will be the turning
point. Using as variable $z= y z_0$, the distance $L$ between the two quarks at radial coordinate
$z_b$ is  given by
\be
 {L\over\beta} ={1\over\beta}\,\int d\sigma= {2\over\beta}\int_1^{y_b} {d y\over y^\prime}=
{2\over\pi} \int_1^{y_b}{1\over\sqrt{y^4/\tilde h + \tilde
\gamma^2/\tilde h -1}} {y\over\sqrt{y^4-1}}\,, \ee with $\tilde h=
h/z_0^4$. In the same variables the action can be written as \be
S_r= {1\over 2\pi}\,{e^{\phi_0}\, \hat t\,\over \tilde \gamma}\, 4
\sqrt{N_c} z_0^2 \int_1^{y_b} {y\,(\tilde
\gamma^2+y^4)\over\sqrt{(y^4+\tilde \gamma^2-\tilde h)(y^4-1)}}\,. \ee

We will deal essentially with two possible different limits.

\begin{itemize}
\item[{\it{(a)}}] In the limit $\tilde \gamma\to \infty$, while keeping $y_b$ finite, one must also send
$\tilde h\to \infty$ in order to keep $L$ finite. Indeed one must
have a finite quotient, $\lim(\tilde \gamma^2/\tilde h)\equiv c >1$.
In this limit, $L$ becomes \be {L\over\beta} \approx  {2\over
\pi}\,{1\over \sqrt{c-1}}\, \int_1^{y_b} {y\over\sqrt{y^4-1}} =
{1\over \pi\sqrt{c-1}} \,{\rm arccosh}(y_b^2)\,, \ee and $S_r$, \be
S_r\approx   {1\over 2\pi}4\sqrt{N_c}\hat t e^{\phi_0}z_0^2
\sqrt{c\over c-1}\, {1\over 2}\,{\rm arccosh}(y_b^2)=
{2\pi\lambda\,\hat t\,\over \beta^2}\sqrt{c}\, L\,. \ee When
practicing the next limit, $y_b\to\infty$, $c$ must go to infinity
too in order to keep $L$ finite. The action $S_r$ becomes infinite
and is renormalized by subtracting the infinite mass quark
contribution associated with the action $S_0$. Before taking
$y_b\to\infty$, $S_0$ is given by \be S_0=  {1\over
2\pi}4\sqrt{N_c}\hat t e^{\phi_0}z_0^2 \, {1\over 2}\,{\rm
arccosh}(y_b^2)= {2\pi\lambda\,\hat t\,\over \beta^2}\sqrt{c-1}\,
L\,. \ee Therefore, again the renormalized action vanishes for the
``{\sl down}'' configuration, \be S_{\rm ren} =
\lim_{c\to\infty}(S_r-S_0) =0\,, \ee and accordingly the jet
quenching parameter is zero.

\item[{\it{(b)}}]
As a second possibility we analyze the case where we first take the
$y_b\to\infty$ limit before taking $\tilde \gamma\to \infty$. Notice
that \bea S_r-S_0 &=&  {4\lambda\,\hat{t} \over
\beta}\,\frac{1}{\tilde \gamma}
 \int_1^{\infty}
y\,\sqrt{\tilde \gamma^2+y^4\over y^4-1}
\left(\sqrt{\tilde \gamma^2+y^4\over y^4+\tilde \gamma^2-\tilde h} -1 \right)
\nonumber \\
&\approx& {2\lambda\,\hat{t} \over \beta}\,\frac{\tilde h}{\tilde
\gamma}\, \,\int_1^{\infty} \, \frac{y}{\sqrt{(\tilde \gamma^2+y^4)(
y^4-1)}} \,, \eea with $\tilde h < \tilde \gamma^2 +1$. Using a
similar expansion for $L$, to first order in $\tilde h/\tilde
\gamma^2$ \be
 {L\over\beta}
\approx {2\over\pi}\sqrt{\tilde h} \int_1^{\infty}{y\over\sqrt{(y^4
+ \tilde \gamma^2)(y^4-1)}}\,, \ee which shows that, at this leading
order, \be S_r-S_0 \approx {4\lambda\,\hat{t} \over
\beta}\frac{1}{2}\,\frac{\pi}{2}\, \frac{\sqrt{\tilde h}}{\tilde
\gamma}{L\over\beta}= {\pi\,\lambda\,\hat{t} \over \beta^2}\,
\frac{1}{\sqrt{c}}\,L\,, \ee which again vanishes when $c\to\infty$
(the limit $c\to\infty$ must be taken when $\tilde \gamma\to\infty$
in order to keep $L$ finite).

\end{itemize}

 \medskip

\subsubsection{The {\sl inner} configuration}

In this case we will make use of the parameter $\tilde \alpha^2= 1 +
{ y_t^4 \over \tilde \gamma^2}$. The distance between the two quarks
is \be {L\over \beta} ={2\over \beta}\, \int_{z_t}^{z_b} {dz\over
z'} = {2\over \pi} \,\tilde \alpha \tilde \gamma \int_{y_t}^{y_b} {
y dy\over \sqrt{(y^4-y_t^4)(y^4-1)}}\,. \ee
 Notice that, as said above, the limit when the turning point touches the horizon, $y_t\to 1$, is not
compatible with keeping $L$ finite. In such case the integral develops a
singularity that should be
canceled with the proper limit for the prefactors, $\lim_{y_t\to 1}(\tilde \alpha \tilde \gamma)\to 0$. In this case
$\tilde \alpha \tilde \gamma =
 {\tilde \alpha y_t^2 \over \sqrt{\tilde \alpha^2- 1}} \to
{\tilde \alpha  \over \sqrt{\tilde \alpha^2- 1}}$ and since $\tilde
\alpha\geq 1$, this limit can never go to zero, and the cancelation
can not occur.

In the configuration $y_t \gg 1$  we obtain \be\label{length-inner}
{L\over \beta} \approx {\tilde \alpha\over \pi \sqrt{\tilde
\alpha^2- 1}}\arccos(\eta^2)\,. \ee Thus each value of $\tilde \alpha$
fixes a maximum length, $ {L\over \beta}\leq {\tilde \alpha\over 2
\sqrt{\tilde \alpha^2- 1}}$.

The action, within the same condition $y_t \gg 1$, is
\be\label{Sinner}
 S_r\approx   {2\lambda \hat t\over \beta} \Big(\sqrt{\tilde \alpha^2-1}\,
\tan({ \pi\,\sqrt{\tilde \alpha^2-1}\,L\over \beta \tilde \alpha}) +
{ \pi \over \beta}{L\over \tilde \alpha}  \Big)\,. \ee We take $\tilde
\gamma, L, y_b$ as the free parameters. Notice that now it is not
possible to keep $v>1$ while having $V=1$. To get the lightlike
situation we need to send $v\to 1$ and $y_b\to \infty$. These limits
can be done now independently. Sending $v\to 1$ from above is
equivalent to $\tilde \gamma\to \infty$. We will examine the
different possibilities.
\begin{itemize}
\item[{\it{(a)}}] We take first the
$\tilde \gamma\to \infty$ limit keeping $y_b$ finite. Since now
\be\label{quotbis} {y_b^4 \over \tilde \gamma^2}={\tilde
\alpha^2-1\over \eta^4}= {\tilde \alpha^2-1\over \Big(\cos({
\pi\,\sqrt{\tilde \alpha^2-1}\,L\over \beta \tilde
\alpha})\Big)^2}\,, \ee we end up with a single possibility for  $
\tilde \alpha$, which is $ \tilde \alpha\to 1$. In addition this
entails $y_t\to y_b$, so the string stays on the brane (``{\sl
short}" string) as long as $y_b$ is finite and we obtain $S_r=
{2\pi\lambda\hat{t} \over \beta^2}\,L$. We can send next $y_b \to
\infty$, but this limit does not affect $S_r$. In addition to that,
as already happened in the $v<1$ case, the subtraction corresponding
to two straight strings from the location of the brane at $y_b$ and
the horizon at $y=1$ becomes infinite when $y_b \to \infty$. Indeed
the subtraction term, \be\label{subtrp} S_0 = {4\lambda\,\hat{t}
\over \beta\,\tilde\gamma}\int_1^{y_b} dy y
\sqrt{\frac{\tilde\gamma^2 + y^4}{y^4-1}} \ee
 behaves, for
$\tilde\gamma\to\infty$ and finite $y_b\gg 1$ as
$S_0 \approx
{2\lambda\,\hat{t} \over \beta}\log(2y_b^2)\,.$
Thus, again, we can not get a finite renormalized action.
\medskip

\item[{\it{(b)}}] Next we can consider to take first the limit $y_b\to\infty$. Sending (\ref{quotbis}) to infinity can be
done essentially in two different ways depending on the actual value of $L$.

\begin{itemize}
\item[{\it{(b.1)}}] Sending $\tilde\alpha\to\infty$ whereas ${2 L\over\beta}\leq 1$.
\smallskip

When ${2 L\over\beta}< 1$, using (\ref{Sinner}), we find in the
large $\tilde\alpha$ limit that the behavior of the action is $S_r =
{2\lambda \hat t\over \beta}\, \tilde \alpha\, \tan({ \pi\,\,L\over
\beta }) + {\cal O}(\frac{1}{\tilde \alpha})$. On the other hand,
the subtraction term $S_0$ in the large $y_b$ limit is $S_0 =
{4\lambda \hat t\over \beta\tilde \gamma} \frac{y_b^2}{2} + {\rm
finite}$. The subtraction cannot cancel the divergences because for
large $\tilde \alpha$ we obtain from (\ref{quotbis}) that
\be\label{ybofalpha} \frac{y_b^2}{\tilde \gamma} =
\frac{\sqrt{\tilde \alpha^2-1}}{\cos({ \pi\,\sqrt{\tilde
\alpha^2-1}\,L\over \beta \tilde \alpha})}\approx \frac{\tilde
\alpha}{\cos(\frac{ \pi\,\,L}{ \beta })} \ee and thus, for large
$\tilde \alpha$ \be S_r-S_0 = {2\lambda \hat t\over \beta} \
\frac{\sin({ \pi\,\,L\over \beta })-1}{\cos({ \pi\,\,L\over \beta
})}\ \tilde \alpha\,, \ee which diverges linearly with $\tilde
\alpha$.

\item[{\it{(b.2)}}] The critical case ${ L\over \beta }= { \pi\over 2 }$ is more subtle. In this
case, the expansion of (\ref{Sinner}) for large $\tilde \alpha$
gives \be S_r = {2\lambda \hat t\over
\beta}\left(\frac{4}{\pi}\,\tilde \alpha^3 - \frac{3}{\pi}\,\tilde
\alpha + {\cal O}(\frac{1}{\tilde \alpha})\right)\,. \ee On the
other hand, using (\ref{ybofalpha}) and (\ref{subtrp}), and
expanding for large $\tilde \alpha$, we get \be S_0= {2\lambda \hat
t\over \beta}\left(\frac{4}{\pi}\,\tilde \alpha^3 -
\frac{3}{\pi}\,\tilde \alpha + {\cal O}(\frac{1}{\tilde
\alpha})\right) + {\rm finite}\,. \ee Thus the $\tilde
\alpha$-divergences cancel out and we end up with a finite
contribution coming form $S_0$, as long as $\tilde\gamma$ is still
kept finite. To isolate this finite contribution it is convenient to
consider the subtraction $S_0 -{\rm divergences}$, \be S_0 -{\rm
divergences} ={2\lambda \hat t\over \beta}
\int^{x_b}_{\frac{1}{\!\!\sqrt{\tilde\gamma}}}\,dx\, x\,
\left({\sqrt{1+x^4}\over \sqrt{x^4-1/\tilde\gamma^2}}-1\right)\,,
\ee where we have used the definition $y=\sqrt{\tilde\gamma}\,x$.

This integration is finite for $x_b\to \infty$ as long as $\tilde\gamma$ is kept finite. But
since the last step of our
whole procedure must be to send
$\tilde\gamma \to \infty$, we see that the integration above develops a logarithmic divergence
in $\tilde\gamma$. Thus it is not possible to obtain a finite renormalized action in the critical
case ${ L\over \beta }= { \pi\over 2 }$.

\item[{\it{(b.3)}}] A second possible limit is
keeping a finite $\tilde\alpha>1$ such that $\cos({ \pi\,\sqrt{\tilde \alpha^2-1}\,L\over
\beta \tilde \alpha})= 0$, which is
${2 L\over\beta}= {\tilde \alpha\over\sqrt{\tilde \alpha^2-1}} > 1$.

Using $y_b=1/\epsilon$, the action becomes, up to an irrelevant
factor ${4\lambda\,\hat{t} \over \beta}$, \be\label{corr} S_r
\approx {1\over \tilde \gamma}
\int_{\tilde\gamma^{1/2}(\tilde\alpha^2-1)^{1/4}}^{1/\epsilon}
{(y^4+\tilde\gamma^2) dy\over
y\sqrt{y^4-\tilde\gamma^2(\tilde\alpha^2-1)}} \approx  {1\over 2
\epsilon^2 \tilde\gamma^2} + {\pi\over 4\sqrt{\tilde\alpha^2-1}}\,.
\ee In order to evaluate the contribution of the straight strings
\be\label{sst} S_0 =  {1\over \tilde\gamma} \int^{1/\epsilon}_1
{\sqrt{\tilde\gamma^2+y^4}\ y\ dy\over \sqrt{y^4-1}}\,, \ee we split
the integral at the point
$\sqrt{\tilde\gamma}(\tilde\alpha^2-1)^{1/4}$. The part above this
point gives \be\label{above} S_0^{above} \approx {1\over
\tilde\gamma} \int^{1/\epsilon}_{\sqrt{\tilde\gamma}
(\tilde\alpha^2-1)^{1/4}} {\sqrt{\tilde\gamma^2+y^4}\ dy\over
y}\approx -{\tilde\alpha\over 2}+ {1\over 2\tilde\gamma \epsilon^2}
+ {1\over 2}\log{\tilde\alpha+1\over \sqrt{\tilde\alpha^2-1}}\,, \ee
canceling the divergence in $S_r$. The remaining piece, below that
point, is \be\label{below} S_0^{below} \approx {1\over \tilde\gamma}
\int_{1}^{\sqrt{\tilde\gamma}(\alpha^2-1)^{1/4}} {\tilde\gamma y\
dy\over \sqrt{y^4-1}}\approx {\log 2\over 2}+{\log\tilde\gamma\over
2}
 +{1\over 2}\log{\sqrt{\tilde\alpha^2-1}}\,,
\ee
which becomes divergent when finally  $\tilde\gamma$ is sent to infinity, thus
again we reach the conclusion that the action diverges.

\end{itemize}
\end{itemize}

\bigskip

Summing up, except for the ``{\sl down}" configurations reaching the
horizon, in all the cases there is no sensible way to obtain a
finite renormalized action for the Wilson loop, in the limit from a
spacelike to a lightlike configuration. The ``{\sl down}"
configurations, on the other hand, yield a vanishing renormalized
action and therefore a zero value for the jet quenching parameter
$\hat q$, thus confirming the findings in \cite{noi}.

\vskip 15pt
\centerline{\bf Acknowledgments}
\vskip 10pt
\noindent

We thank G. Bertoldi, F. Bigazzi, R. Casero, R. Emparan, C. Nunez
and A. Paredes for useful discussions and comments. This work is
partially supported by the European Commission contracts
MRTN-CT-2004-005104, MRTN-CT-2004-503369, CYT FPA 2004-04582-C02-01,
CIRIT GC 2001SGR-00065, MEIF-CT-2006-024173. The work of P.T. is
partially supported by MEC PR2006-0495 and by the RyC program.
A.L.C. thanks the Galileo Galilei Institute for Theoretical Physics
for the hospitality and the INFN for partial support during the
completion of this work.


\end{document}